\documentstyle[sprocl,epsf]{article}

\begin{document}

\title{GLUONS IN THE LATTICE SU(2) CLASSICAL GAUGE FIELD}

\author{YING CHEN, JI-MIN
WU, BING HE}

\address{Institute of High Energy Physics, Chinese Academy of Sciences,\\
P.O.Box 918-4, Beijing 100039, P. R. China\\
E-mail: cheny@hptc5.ihep.ac.cn} 

\maketitle\abstracts{The SU(2) gluonic correlation functions, glueball
effective masses in the $J^{P}=0^{+}$, $2^{+}$ and $0^{-}$ channels were
calculated from the lattice classical gauge configurations which were
obtained by smoothing the thermal gauge configurations through the
improved cooling method. The instanton-induced attractive force in the
$0^{+}$ channel and the repulsive force in the $0^-$ channel are confirmed
in the Monte Carlo simulation. There is evidence that the instanton vacuum
contribution to the $0^+$ glueball mass is significant.}    

\section{Introduction} QCD predicts there exist glueballs. The structure
and formation mechanism of glueballs are so extricated that it must be
explored nonperturbatively. Extensive Monte Carlo simulations have been
performed and a few statements seem to be firmly established in the
glueball sector\cite{s30}: i) the scalar glueball is the lightest with the
mass in the 1.6-1.8 GeV range; ii) the tensor glueball and the
pseudoscalar one are significantly heavier than the scalar one with the
mass ratio $m_{2^{++}}/m_{0^{++}} \simeq 1.4$ and
$m_{0^{-+}}/m_{0^{++}}\simeq 1.5-1.8$, respectively\cite{s31}; iii) The
scalar is much smaller than other glueballs($r_{0^{++}}\simeq 0.2fm$,
$r_{2^{++}}\simeq 0.8fm$\cite{s32}). From these statements one can find
that glueballs are much heavier than typical quark-model hadrons and the
size difference betweensi the scalar and the tensor glueball is much
larger than that of the similar mesons( a similar measurement for the
$\pi$ and $\rho$ mesons gives $0.32fm$ and $0.45fm$). This indicates that
the spin-dependent forces between gluons are stronger than that between
quarks.  \par The information about the gluonic interactions can be
obtained by studying the correlation functions of gluonic operators with
the relevant quantum numbers. Sch\"{a}fer and Shuryak\cite{s4} calculated
the correlation functions and the Bethe Salpeter amplitudes in an
instanton-based model of the QCD vacuum. Their results show that
instantons lead to a strong attractive force in the $J^{PC}=0^{++}$
channel, thus the scalar glueball is much smaller than other glueballs. In
the $0^{-+}$ channel the corresponding force is repulsive, and in the
$2^{++}$ case it is absent. \par Recently the cooling method, a smoothing
algorithm for the gauge configurations, has been developed to extract
classical contents( such as instantons) of QCD vacuum on the lattice. This
motivates us to study the relationship of instantons and glueballs by the
lattice simulation. The goal of this work is to study the instanton
effects in the glueball masses and gluonic correlation functions in the
lattice SU(2) gauge theory. The tadpole improved Symanzik's
action\cite{s6} is used on an anisotropic lattice to perform the Monte
Carlo simulation as accurate as possible on coarse lattices. The smearing
and fuzzy scheme\cite{s71} are applied to the glueball operators to
increase the signal-to-noise ratio. These schemes have been verified to be
successful in Morningstar and Peardon's calculation\cite{s7} and in our
work. We employ the improved cooling to extract the instanton
configurations. The gluonic correlators are measured during the cooling
procedure and the glueball masses are extracted by the exponential
fall-off of these correlators with respect to the Euclidean time. To
investigate the spin dependence of the instanton effects, the behaviors of
the gluonic vacuum correlation functions are also explored in this work.
\par The simulation details, the results and the discussion are given in
section II. Section III is the conclusion and summary.

\section{Lattice Calculation}
 
If the lattice action is properly improved, the costs of the Monte Carlo
simulation will be greatly reduced and the numerical study can be
performed on much coarser lattice system. In this work, the gluonic
correlation functions and the masses of the lowest-lying glueballs in the
SU(2) gauge theory are calculated from uncooled and cooled configurations.
The simulations are performed on an $8^3\times 24$ anisotropic lattice by
using tadpole-improved Symanzik's action\cite{s6,s11}. The lattice spacing
in the temporal direction is smaller than that in the spatial direction
with the anisotropy ratio $\xi=a_s/a_t=3$, so that the physically useful
signals can be obtained before they would be undermined by the statistical
fluctuations. The form of the lattice gauge action we use is expressed as
\begin{eqnarray} S_1 &=& \beta
\sum\limits_{x,s>s'}\left[\frac{5}{3}\frac{P_{ss'}}{\xi
u_s^4}-\frac{1}{12}\frac{R_{ss'}}{\xi u_s^6}
-\frac{1}{12}\frac{R_{s's}}{\xi u_s^6}\right]\\\nonumber &+&\beta
\sum\limits_{x,s}\left[\frac{4}{3}\frac{P_{st}\xi}{u_s^2 u_t^2}
-\frac{1}{12}\frac{R_{st} \xi}{u_s^4 u_t^2}\right], \end{eqnarray} where
$\beta=\frac{4}{g^2}$, $g$ is the gauge coupling. $P_{\mu \nu}$ is the
plaquette operator and $R_{\mu \nu}$ $2\times 1$ rectangular operator,
$u_s$ and $u_t$ are tadpole improvement parameters. $u_s$ is defined by
$u_s=<(1/2)TrW_{sp}>^{1/4}$, while $u_t$ is set to be 1 because of
$u_t=1-O(\xi^{-2})$ ($\xi$ is much bigger than one). \par In the continuum
theory, the gluonic operators with quantum numbers $0^{+}$,
$0^{-}$,$2^{+}$ are defined by the field strength squared, the topological
charge density, and the energy density, respectively, 

\begin{eqnarray}
O_S=(g G_{\mu\nu}^a)^2, O_P=\frac{1}{2}\epsilon_{\mu\nu\rho\sigma}g^2
G_{\mu\nu}G_{\rho\sigma},\\\nonumber O_T=\frac{1}{4}(g G_{\mu\nu}^a)^2-g^2
G_{0\alpha}^aG_{0\alpha}^a. 
\end{eqnarray}
 
The correlation functions $C(x)$ for the Euclidean separation $x$ are
defined as 
\begin{equation} C_{\Gamma}(x)=\langle
0|O_{\Gamma}(x)O_{\Gamma}(0)|0\rangle. 
\end{equation} 

The glueball spectra are obtained by the spatial integral of two points
functions of operators with definite $J^{PC}$ quantum numbers, such as
\begin{eqnarray} C_{\Gamma}(t)&=&\int d \vec{x} \langle
0|O_{\Gamma}(x)O_{\Gamma}(0)|0\rangle\\\nonumber
&=&\sum\limits_{n_{\Gamma}} \frac{1}{2m_{n_{\Gamma}}}|\langle
0|O_{\Gamma}|n_{\Gamma},\vec{p}=0\rangle|^2 e^{-m_{n_{\Gamma}}
|t|}\\\nonumber 
&\propto& |\langle
0|O_{\Gamma}|0_{\Gamma},\vec{p}=0\rangle|^2 e^{-m_{0_{\Gamma}}|t|}
  (t\rightarrow \infty). \end{eqnarray} The lattice simulation of masses
of lowest-lying hadrons is well established. The lattice version of
$O_{\Gamma}$ are defined as \cite{s9} \begin{eqnarray}
O_{0^+}(t)&=&\sum\limits_{\vec{x}}[P_{12}(\vec{x},t)+P_{23}(\vec{x},t)
+P_{31}(\vec{x},t)],\\
O_{2^+}(t)&=&\sum\limits_{\vec{x}}[P_{12}(\vec{x},t)-P_{13}(\vec{x},t)].
\end{eqnarray} The corresponding glueball masses are extracted from the
exponential fallof of the lattice correlation functions. \begin{equation}
m_{\Gamma}(t)=\frac{1}{a_t}ln\left(\frac{C(t)}{C(t-a_t)}\right)
\end{equation} We first calculate the lowest-lying glueball masses on the
uncooled configurations at $\beta=1.0,1.1,1.2$ through the standard
procedure, where the fuzzy and smearing method\cite{s71,s7} are used to
increase the signal-to-noise ratio. The configurations are thermalized
through 5000 Monte Carlo sweeps( each sweep is composed of four heatbath
sweeps and one micro-canonical iteration) and the measurements are
performed on 10000 configurations. We analyze the statistical error by
dividing the 10000 configuration into 20 bins. The physical scale is set
by calculating
the heavy quark potential. With the value of the string tension
$\sqrt{\sigma}=440 MeV$, $a_s$ takes
the value 0.277$fm$, 0.233$fm$, and 0.183$fm$\cite{s8} respectively for
$\beta=1.0,1.1,1.2$. The glueball masses at $\beta=1.0$ are extracted to
be
  $$m_{0^+}=1634\pm 60 MeV$$
   $$m_{2^+}=2305\pm 72 MeV$$
 $$m_{0^-}=2574\pm 224 MeV$$ Our results are in good agreement with that
of other work\cite{s8}. \par To explore the role of instantons in the
gluonic correlation functions and glueball masses, the cooling method is
used to get the instanton configurations. The link-updating of the cooling
iteration is defined as below $$
        U_{\mu}\rightarrow U_{\mu}'=c\Sigma_{\mu}^{+}, $$ where $c$ is the
normalization factor so that $U_{\mu}'$ is an element of SU(2) group.
$\Sigma_{\mu}$ is defined by the local action $$
s(U_{\mu})=1-\frac{1}{2}TrU_{\mu}\sum\limits_{\nu\neq \mu}(staples)\equiv
1-\frac{1}{2}TrU_{\mu}\Sigma_{\mu}, $$ This transformation justifies that
the local action will be minimized. We take the measurement on a sample of
500 configurations generated and cooled by the lattice action $S_1$. The
gluonic correlators are measured every five cooling sweeps and the spectra
are extracted by the same approach as that was used in the uncooled
case.In order to avoid the auto-correlation, every two adjacent
configuration are separated by 20 Monte Carlo sweeps. Each configuration
is cooled by 100 steps. The cooling procedure is monitored by calculating
topological charges at each cooling step. The lattice charge density
operator($Q_{cont.}(x)=\frac{1}{16\pi
^2}TrF_{\mu\nu}(x)\tilde{F}_{\mu\nu}(x)$ in continuum theory) used in this
work is the improved version, \begin{eqnarray} Q_L(x)&=&-\frac{1}{512
\pi^2}\left(\frac{5}{3}Q_P(x)-\frac{1}{6}Q_S(x)\right)\\\nonumber &=&a^4
(Q_{cont.}(x)+O(a^4)), \end{eqnarray} where \begin{eqnarray}
Q_P(x)&=&\sum\limits_{\mu\nu\rho\sigma=\pm 1}^{\pm 4}
\tilde{\epsilon}_{\mu\nu\rho\sigma}Tr\left(P_{\mu\nu}(x)
P_{\rho\sigma}(x)\right) \\\nonumber
Q_S(x)&=&\sum\limits_{\mu\nu\rho\sigma=\pm 1}^{\pm
4}\tilde{\epsilon}_{\mu\nu\rho\sigma}Tr\left[\left(S_{\mu\nu}^{1\times
2}(x)+S_{\mu\nu}^{2\times 1}(x)\right)\left(S_{\rho\sigma}^{1\times
2}(x)+S_{\rho\sigma}^{2\times 1}(x)\right)\right].\\\nonumber
\end{eqnarray} Here $\tilde{\epsilon}_{\mu\nu\rho\sigma}$ is the standard
Levi-Civita tensor for positive directions while for negative ones the
relation
$\tilde{\epsilon}_{\mu\nu\rho\sigma}=-\tilde{\epsilon}_{-\mu\nu\rho\sigma}$
holds. $P_{\mu\nu}$ is the plaquette operator in the $\mu-\nu$ plane,
while $S_{\mu\nu}^{i\times j}$ is $i\times j$ rectangular operator. We
find that after several cooling sweeps the charges will reach meta-stable
plateaus with approximate integer values. \par Fig. 1, 2, 3 show the
correlation functions of the three channels. In each channel, the
correlation function is plotted at 10, 25, 50 and 100 cooling steps. The
glueball "masses" are extracted in the same way as in the uncooled case
and are plotted in Fig. 4, 5, 6. 

\begin{figure}[t] 
\label{corr0}
\epsfysize=2in 
\hspace{3.5cm}  
\epsffile{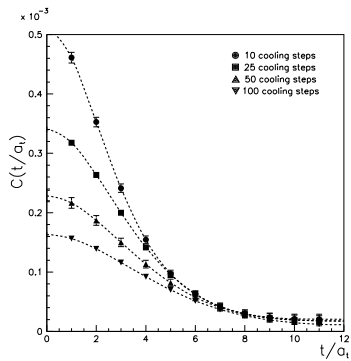}   
\caption{ The correlation function
C(t) in $0^{+}$ channel at various cooling steps are illustrated. The
dots, the rectangulars, the triangles and the upside-down triangles
correspond to the cooling steps 10, 25, 50 and 100, respectively. }
\end{figure} 

\begin{figure}[t] 
\epsfysize=2in
\hspace{3.5cm}
\epsffile{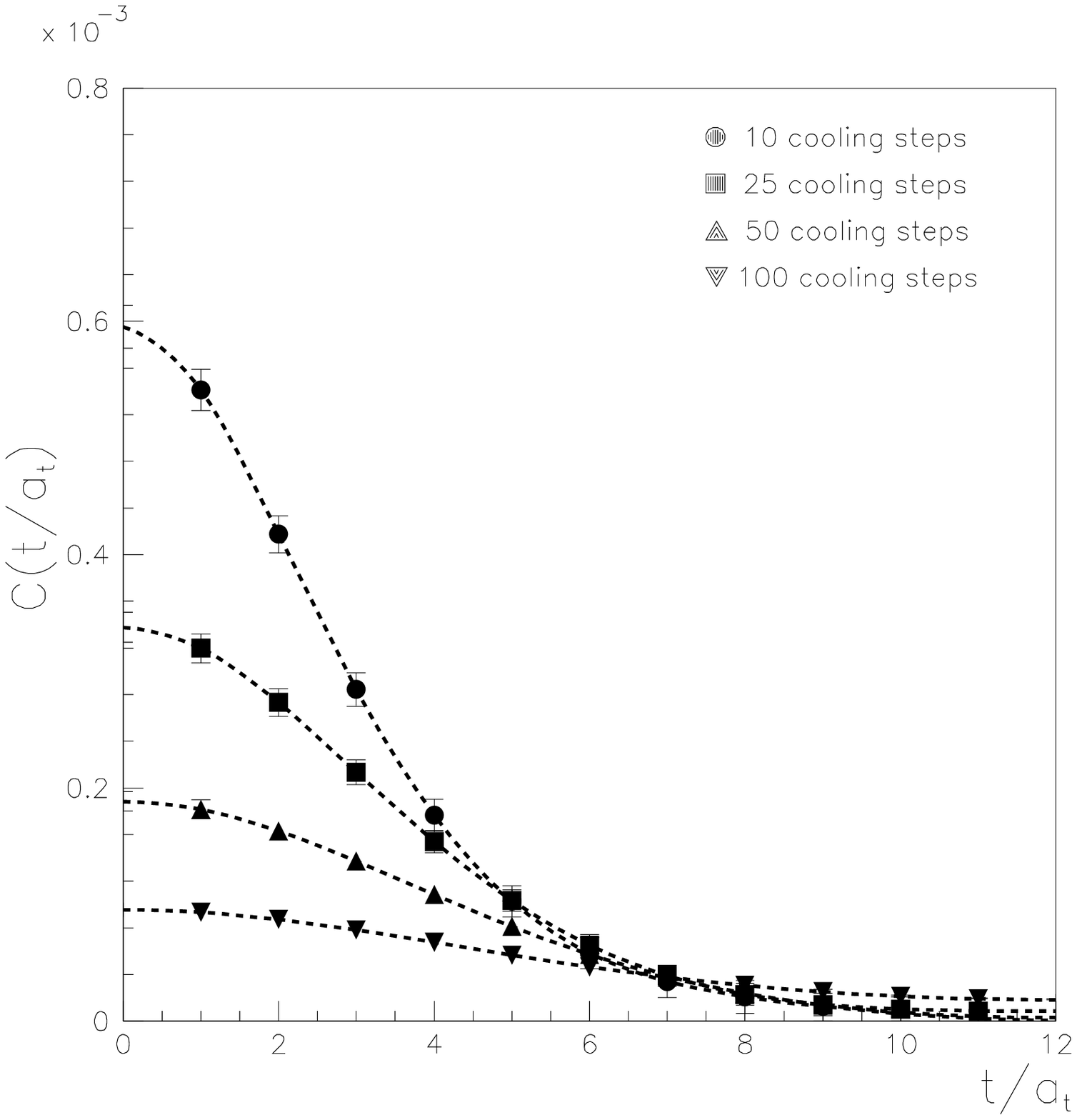}      
 \caption{ The
correlation functions C(t) in $2^{+}$ channel at various cooling steps are
illustrated. The dots, the rectangulars, the triangles and the upside-down
triangles correspond to the cooling steps 10, 25, 50 and 100,
respectively. } 
\end{figure} 

\begin{figure}[t]
\epsfysize=2in 
\hspace{3.5cm}  
\epsffile{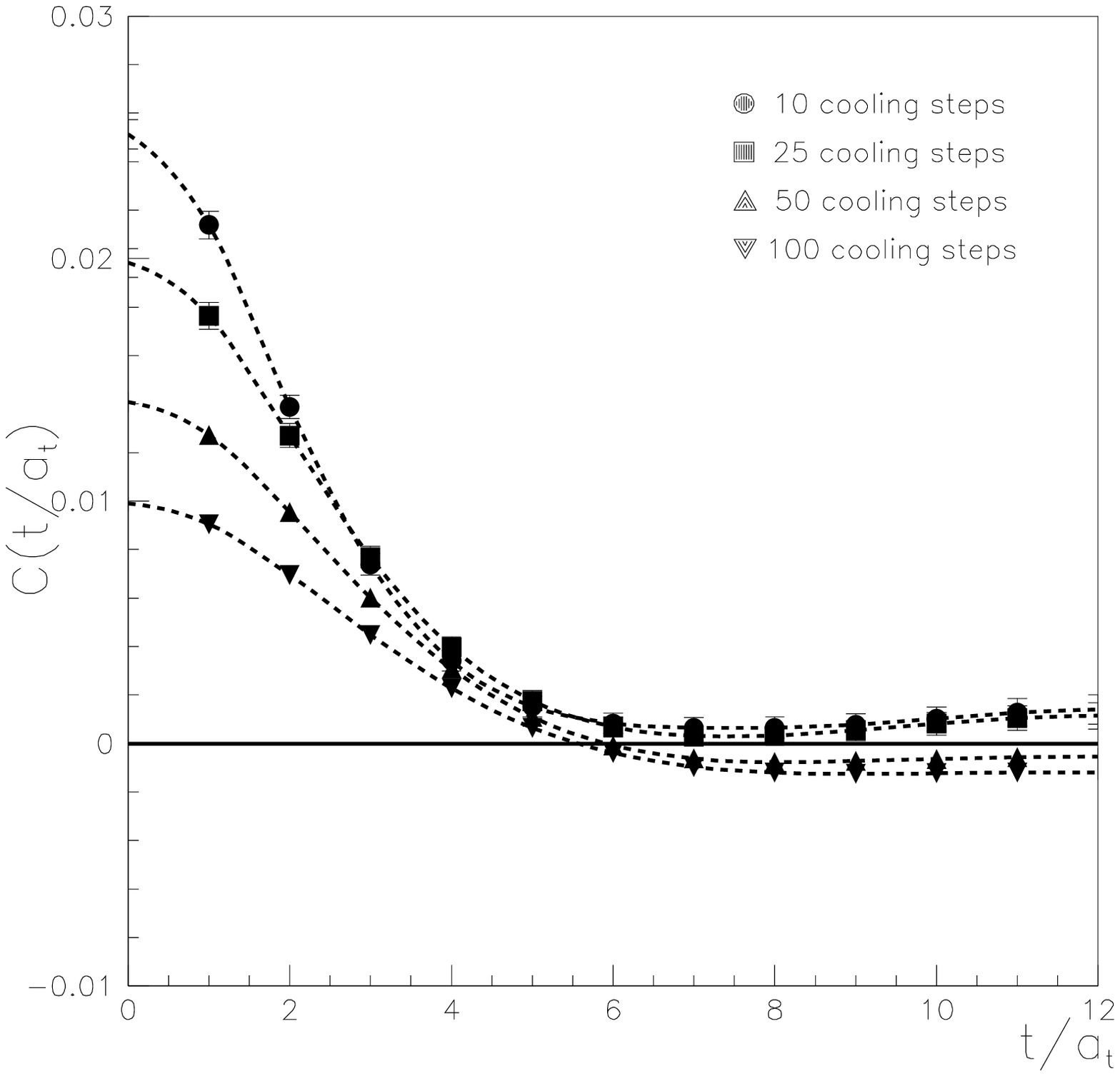}
\caption{ The correlation functions C(t) in $0^{-}$ channel at various
cooling steps are illustrated. The dots,
the rectangulars, the triangles and the upside-down triangles correspond
to the cooling steps 10, 25, 50 and 100, respectively.} 
\end{figure}

\begin{figure}[t]
\epsfysize=2in 
\hspace{3.5cm}
\epsffile{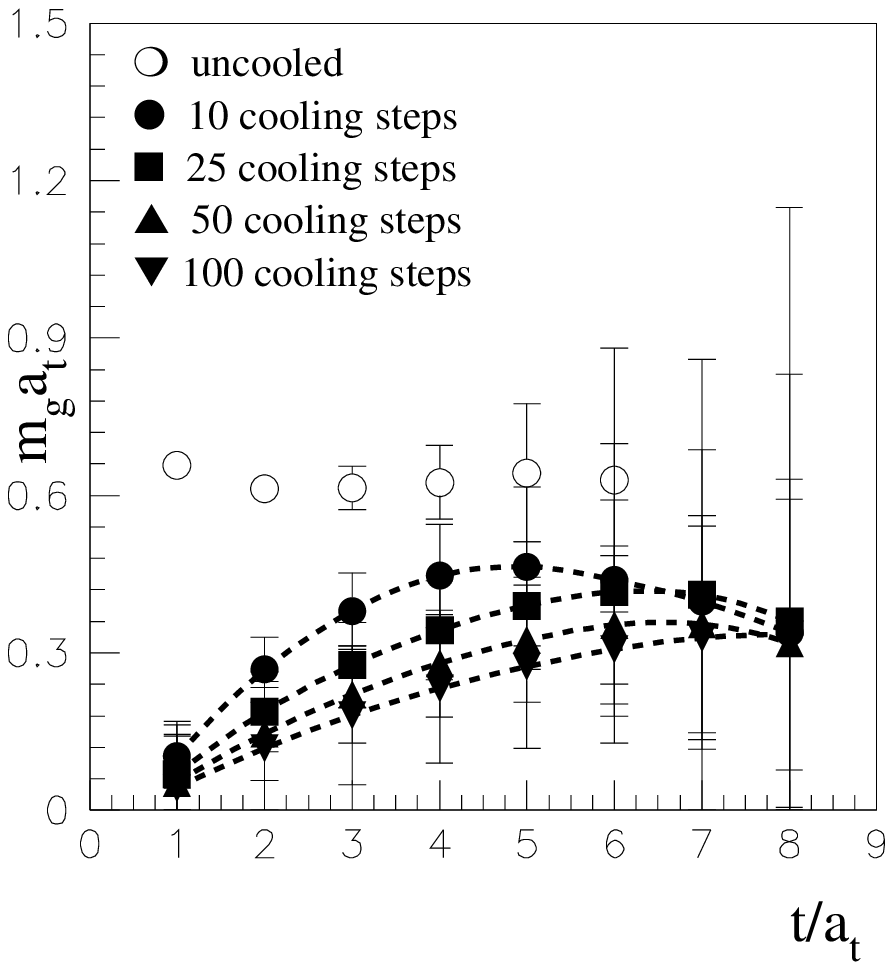}
\caption{ The $0^{+}$ channel 'glueball masses' are
extracted from the exponential decay of the correlators in cooled case.
The masses vs. Euclidean time at various cooling steps are plotted. }
\end{figure} 
\begin{figure}[p]
\epsfysize=2in 
\hspace{3.5cm}
\epsffile{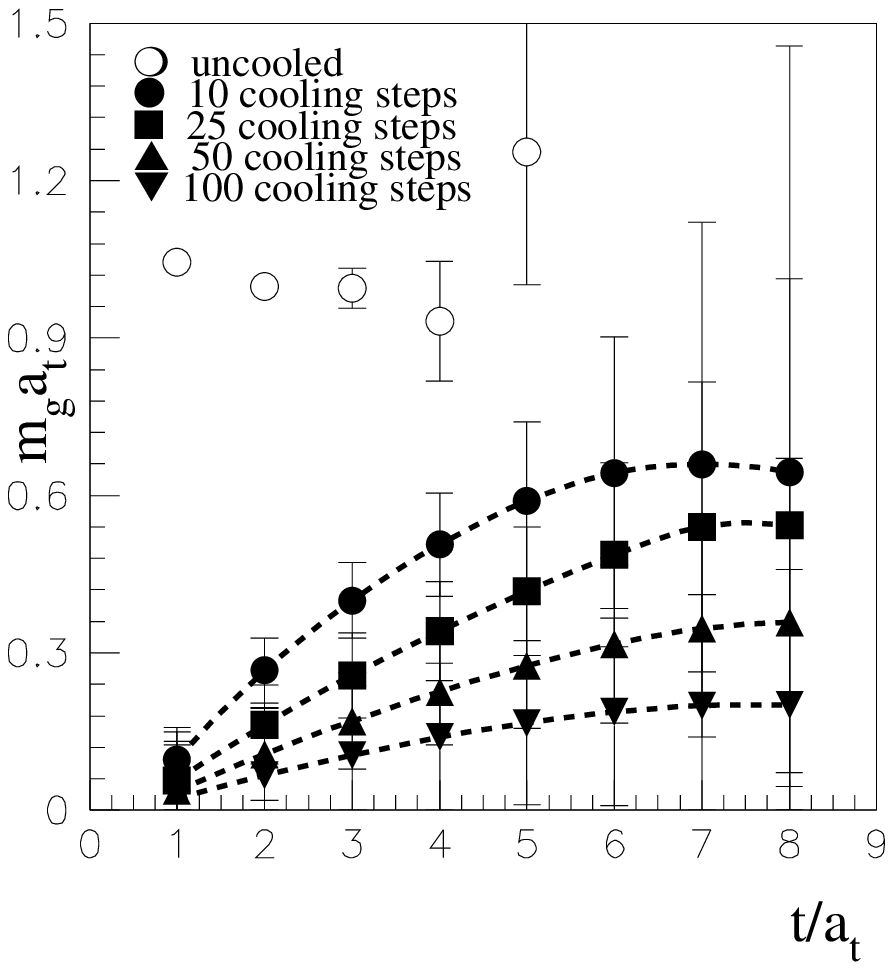} 
\caption{ The $2^{+}$ channel 'glueball masses' are
extracted from the exponential decay of the correlators in cooled case.
The masses vs. Euclidean time at various cooling steps are plotted. }
\end{figure} 
\begin{figure}
\epsfysize=2in 
\hspace{3.5cm}   
\epsffile{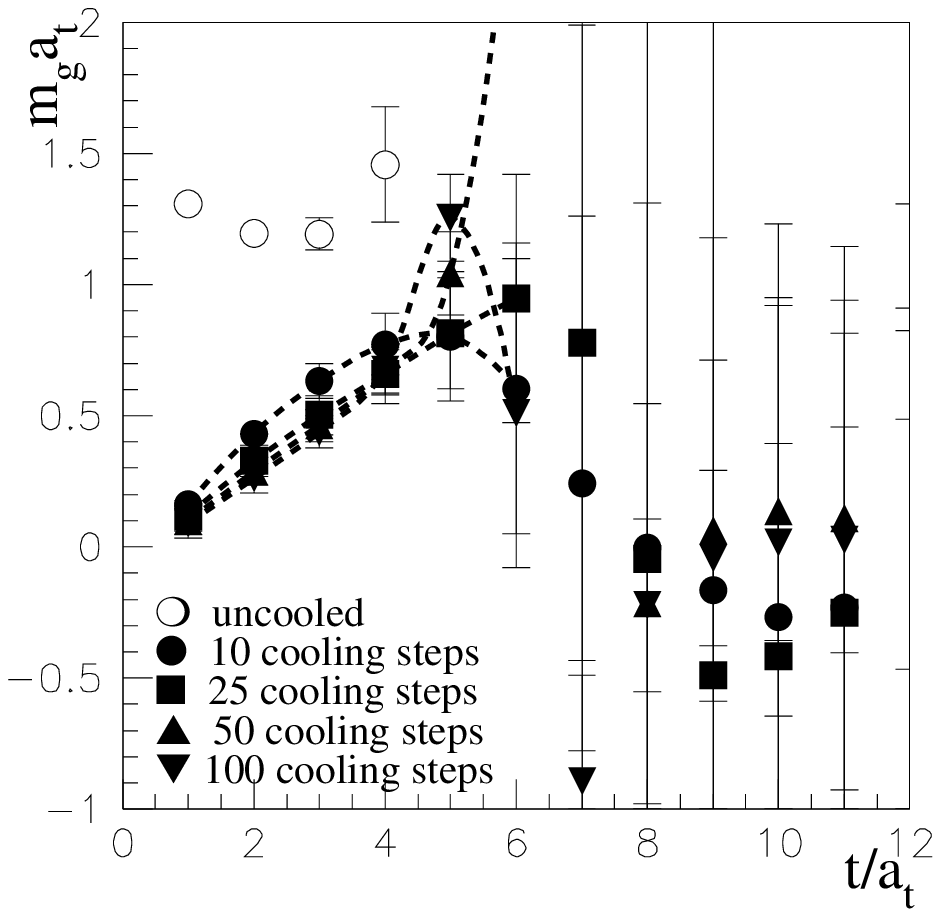} 
\caption{ The $0^{-}$ channel 'glueball masses' are
extracted from the exponential decay of the correlators in cooled case.
The masses vs. Euclidean time at various cooling steps are plotted. No
plateaus appear. The mass information can not be obtained in this channel.
} 
\end{figure} 
\par Shuryak and Sch\"{a}fer has calculated the gluonic
correlation functions in the single-instanton background field. Taking the
free gluon propagator into consideration, the correlators are expressed as
\cite{s4}, \begin{eqnarray} C_{S}(x)&=&\frac{384g^4}{\pi^4 x^8}+n\rho^4
C_{inst}(x),\\ C_{P}(x)&=&-\frac{384g^4}{\pi^4 x^8}+n\rho^4 C_{inst}(x),\\
C_{T}(x)&=&\frac{24g^4}{\pi^4 x^8}, \end{eqnarray} where $g$ is the
running coupling constant and $C_{inst}(x)$ is the instanton contribution.
$C_{inst}(x)$ is definitely positive and its detailed expression is
abbreviated here. The signs in Eqn.(10) and Eqn.(11) are impressive: the
two contributions reinforce in the scalar channel(attractive) but cancel
to some extent in the pseudoscalar channel(repulsive). There is no effect
in the tensor channel. \par Our results shown in the figures are
qualitatively in agreement with these comments. At small $t$, the
amplitudes of the correlation functions decrease rapidly with the increase
of the cooling steps, so do the extracted masses. In the intermediate and
large $t$ range, the correlation functions are less sensitive to the
cooling. The "masses" take out short plateaus in the $0^+$ channel and the
$2^+$ channel in the intermediate $t$. These are the results of that the
ultraviolet components of the gauge fields are removed by the cooling but
their topological structure is kept more stable. Small instantons are
smoothed out by the cooling iterations(we monitor the cooling procedure
and find that only the instantons with sizes greater than $2 a_s$
survive), so the correlation functions and the "masses" at small $t$ may
be underestimated. The fact that in the $2^+$ channel the correlation
function tends to be flat(Fig. 2) and the mass-plateau decreases rapidly
(Fig. 5) imply that there is no instanton effect in this channel. In
contrast to this, the fall-off the correlation functions with $t$ in the
$0^{+}$ and $0^{-}$ channels are obvious even after 100 cooling sweeps.
The height of the mass plateau in the $0^{+}$ channel is less sensitive to
the cooling steps in the intermediate $t$ range(Fig. 4). This means the
contribution of instantons to the glueball mass in the $0^+$ channel is
significant. As illustrated in Fig. 3, the correlation function in the
$0^{-}$ channel even changes sign at $t=5 a_t-8 a_t$ after 50 cooling
steps. This is an evidence of that the gluonic interaction induced by
instantons is repulsive in the pseudoscalar channel. The mass information
can not be clearly extracted in this channel (Fig. 6). 
\section{Conclusion} 
We simulated the SU(2) glueball masses in the cooled and
cooled gauge configurations. The results in the uncooled case are in
agreement with the existing work. In the cooled case there is evidence of
the instanton-induced attractive and repulsive interactions in the $0^+$
channel and the $0^-$ channel, respectively, which are predicted by
Shuryak and Sch\"{a}fer based on the instanton liquid models. The
contribution of instantons to the $0^+$ glueball mass is significant.
\section*{Acknowledgement} The
authors thank Dr. Jianbo Zhang and Dr. Daren Ji for their kindly providing
us part of the Fortran code. Some of the calculations were performed on
the Dawning-2000I parallel computer of the National Research Center for
Intelligent Computing System. Y. Chen thanks Prof. Mingfa Zhu, Prof.
Xiangzhen Qiao, Dr. Shilen Yao, and Prof. Jincai Shi for their kind help.
 \end{document}